# Investigation of phytochemicals, spectral properties, anticancer, antidiabetic, and antimicrobial activities of chosen Ayurvedic remedies.

**T. H. MOHAMED AHADU SHAREEF [A]\*, IRFAN NAVABSHAN [B] M MOHAMED DIVAN MASOOD [C] T. ESWARA YUVARAJ [D] A. SHERIF [E]**

[1][a]\*P.G. and Research Department of Chemistry, The New College (Autonomous), Affiliated to the University of Madras, Chennai, Tamil Nadu, India - 600 014.

[b]Crescent School of Pharmacy, B.S. Abdur Rahman Crescent Institute of Science & Technology, Chennai, Tamil Nadu, India - 600 048.

[c] School of Computer Applications, B.S. Abdur Rahman Crescent Institute of Science & Technology, Chennai, Tamil Nadu, India - 600 048.

[D &E]P.G. and Research Department of Chemistry, The New College (Autonomous), Affiliated to the University of Madras, Chennai, Tamil Nadu, India - 600 014.

**\*Corresponding author:** Dr. T.H. Mohamed Ahadu Shareef**,** Assistant Professor of Chemistry**,** P.G. and Research Department of Chemistry**,** The New College (Autonomous), Chennai – 600 014, Tamil Nadu, INDIA.

Mail Id: jasshaali@gmail.com; Mobile: 7904617184 & 9790405029

## ABSTRACT

This study examines the phytochemical characteristics of Ayurvedic products. An analysis was performed on Kottakkal Ayurveda Triphala (T), Kottakkal Ayurveda Hinguvachadi Churnam (H), and Kottakkal Ayurveda Jirakadyarishtam (J) using GC-MS and LC-MS techniques to determine their bioactive constituents, while also assessing their antimicrobial, docking, anticancer, and anti-diabetic activities. The GC-MS analysis identified 30, 45, and 8 chemical components in Kottakkal Ayurveda Triphala (T), Kottakkal Ayurveda Hinguvachadi Churnam (H), and Kottakkal Ayurveda Jirakadyarishtam (J), respectively. The LC-MS analysis produced 15, 20, and 16 peaks for Kottakkal Ayurveda Triphala (T), Kottakkal Ayurveda Hinguvachadi Churnam (H), and Kottakkal Ayurveda Jirakadyarishtam (J), with m/z values of 982, 981, 972, and 933; 987, 985, 974, and 945; and 969, 965, 951, and 941, respectively, confirming their precision. Moreover, characterization of the Ayurvedic products was carried out using FT-IR, UV-vis, and $^1$H-NMR spectroscopy to identify significant functional groups and chemical substances. Kottakkal Ayurveda Triphala (T) was evaluated for antibacterial activity against Gram-positive bacteria (Streptococcus pneumoniae and Staphylococcus aureus) along with Gram-negative bacteria (Escherichia coli and Klebsiella pneumoniae), yielding a P value of 0.0650 (P < 0.0001). Both Kottakkal Ayurveda Hinguvachadi Churnam (H) and Kottakkal Ayurveda Jirakadyarishtam (J) were subjected to analysis for their effectiveness against Aspergillus niger and Aspergillus fumigatus, also revealing a P value within the acceptable range of 0.0650 (P < 0.0001). The anti-diabetic properties of Kottakkal Ayurveda Triphala (T) were assessed using the α-glucosidase inhibitory method, which exhibited a significant inhibitory effect on α-glucosidase, resulting in an average P value of 0.001 (P < 0.0001). Molecular docking evaluations were conducted on three bioactive compounds from the Ayurvedic preparations to examine their interactions with various amino acids. The amino acid at the active site of 6ULM (human cadherin) shows spatial interaction with the 8Ta molecule, which binds to the same amino acid on the surface. The bioactive compound 8Ta from Kottakkal Ayurveda Triphala (T) demonstrated a docking score and calculated MMGBSA dG bind values against 6ULM of -10.73 and -59.52 kcal/mol, respectively. These docking scores and MMGBSA dG bind values were utilized to assess in vitro anticancer activity versus the human colorectal carcinoma cell line (HCT 116) utilizing the MTT assay technique. Extracts from the three Ayurvedic formulations, namely Kottakkal Ayurveda Triphala (T), Kottakkal Ayurveda Hinguvachadi Churnam (H), and Kottakkal Ayurveda Jirakadyarishtam (J) at a concentration of 5 (µg/ml), reduced the development of the human colorectal carcinoma cell line (HCT 116) by 33.42%, 28.06%, and 26.79%, respectively. Therefore, this research intends to enhance the pharmacological knowledge for chemists interested in effectively using antimicrobial, anti-diabetic, and anticancer agents.

1. Introduction

Ayurveda, a term derived from Sanskrit, translates to "science of life" and has its roots in India, dating back around 5000 years. The practice of Ayurvedic medicine is not a recent development; it has been in use for over 3000 years, emphasizing the balance between the body, mind, and spirit. The earliest significant texts on Ayurveda, such as the Susuruta Sangithai and Charaka Sangithai from the Vedic era, showcase how Ayurvedic practitioners devised remedies and surgical techniques to treat various illnesses, establishing a long-standing tradition that remains one of the most influential medical systems in South Asia (Ling, Li, Zhu, Song, & Zhang, 2024). Western medicine recognizes that Ayurvedic practices are not replacements for conventional medicine but rather serve as complementary approaches, requiring strict adherence to guidelines to maximize benefits. It is widely believed that Ayurveda can address any ailment. Ayurvedic treatments aim not only to heal diseases but also to prevent their recurrence once they are resolved (Irfan, Kwak, Han,Hyun, & Rhee, 2020). These treatments can be categorized as either internal or external. Some Ayurvedic remedies are safe for anyone to take at any time, while others should only be used under the guidance of a qualified Ayurvedic practitioner. If deemed necessary, Ayurvedic medicines can be safely taken alongside allopathic treatments without adverse effects. In contemporary times, traditional medicinal practices such as Ayurveda, Yoga, Naturopathy, Unani, and Siddha have garnered global recognition through generations of dedicated practitioners (Ng, Bun, Zhao, & Zhong, 2023). Kerala, a state in India, is renowned for its Ayurvedic heritage, with its residents long believing in this system of healing. Numerous Ayurvedic treatment canters address a variety of health issues, including boosting immunity, managing mental health, alleviating anxiety, pain relief, weight reduction, skincare, sleep disturbances, psoriasis, eye health, arthritis, stroke recovery, sciatica, gastrointestinal issues, and many other conditions (Naseri et al., 2022). According to Ayurveda, there is a significant relationship between the body and the earth, as well as the medicinal plants that grow from it. If the synergy between the body and the earth is disrupted, this holistic medicine, which operates from within, cannot be effective (Liu, Lu, Hu, & Fan, 2020). While we have investigated a handful of medicinal plants, our research has not covered a vast array of them. If Ayurveda is integrated into our daily lives and society, individuals could live in a divine manner. As per the WHO, it was projected that by 2022, 20 million people were affected by cancer, with 9.7 million fatalities resulting from the disease. The statistics indicate that 1 in 9 men, 1 in 12 women, and 1 in 5 individuals on cancer medication fall victim to cancer, with a total of 53.5 million people impacted over the past five years. Numerous cancer research organizations are striving to discover an effective treatment, but despite many efforts, a definitive drug for cancer remains elusive (Wu & Yang, 2020). Additionally, the WHO reported that before reaching 70 years of age, diabetes accounted for 1.6 million deaths, representing 47% of fatalities, with kidney disease causing another 11%, translating to 530,000 deaths, and cardiovascular diseases linked to elevated blood glucose levels. By 2022, a staggering 63% of adults diagnosed with diabetes were not on any medication for their condition (Elekofehinti, Onunkun, & Olaleye, 2020). Likewise, while there is currently no permanent solution for diabetes, temporary treatments are offered based on the severity of the illness. Ayurvedic treatments effectively address various conditions, including anxiety, asthma, arthritis, digestive issues, eczema, hypertension, high cholesterol, rheumatoid arthritis, and stress (Yusupu & Aikemu, 2016). Kottakkal Ayurveda Triphala (T), composed of haritaki, bibhitaki, and amalaki, aids in promoting digestive health, enhancing immunity, and facilitating healthy bowel movements. Kottakkal Ayurveda Hinguvachadi Churnam (H), which contains haritaki, vacha, and hing, is beneficial for treating abdominal colic, loss of appetite, indigestion, and spasms in the intestines and uterus (Hu et al., 2020). The key ingredients in Kottakkal Ayurveda Jirakadyarishtam (J)—cumin seeds, dhataki, ginger, nutmeg, cinnamon, clove, and nagkesar—are valuable for managing dyspepsia and providing post-natal care (Zaman et al., 2021; Hou et al., 2024). Despite their potential, studies assessing the phytochemical (GC-MS and LC-MS), spectroscopic (FT-IR, UV-Vis, and $^1$H-NMR), molecular docking, antimicrobial, anti-diabetes, and anti-cancer properties of Kottakkal Ayurveda Triphala (T), Kottakkal Ayurveda Hinguvachadi Churnam (H), and Kottakkal Ayurveda Jirakadyarishtam (J) have not yet been published. Based on the preceding information and existing research, the three Ayurvedic formulations—Kottakkal Ayurveda Triphala (T), Kottakkal Ayurveda Hinguvachadi Churnam (H), and Kottakkal Ayurveda Jirakadyarishtam (J)—were selected for studies involving phytochemistry (GC-MS and LC-MS), spectroscopy (FT-IR, UV-Vis, and $^1$H-NMR), molecular docking, and investigations into their antimicrobial, anti-diabetic, and anti-cancer properties (Defossez et al., 2021; Zhang, Lu, Liu, & Tang, 2021; Zhang et al., 2021a).

2. Experimental

All chemical reagents used in this research were of high purity quality. FTIR spectra were acquired using KBr pellets on a Shimadzu FTIR-8400 spectrophotometer across a range of 200-4000 cm$^{-1}$. The UV-vis absorption spectra were collected using a Shimadzu UV–vis 1600 A spectrophotometer with a quartz cell within a wavelength range of 200–1500 nm. The $^1$H NMR spectra of the polyherbal ayurvedic formulations were evaluated on a JEOL [Delta V5.3.2] NMR spectrometer model JNM-ECZ400S/L1, utilizing TMS as the internal standard (Shen, He, & Shi, 2021). GC-MS and LC-MS analyses were executed at the Indian Institute of Technology (IIT), Madras, Tamil Nadu. GC-MS analysis of the three ayurvedic products— Kottakkal Ayurveda Triphala (T), Kottakkal Ayurveda Hinguvachadi Churnam (H), and Kottakkal Ayurveda Jirakadyarishtam (J)—was conducted using an Agilent 8890 (GC Model - Agilent 5977 MSD) (Omollo et al., 2024). The parameters set for the analysis included a flow rate of 1.2 mL/min; Q-TOF; a syringe size of 10 μL; hold times of 3/5 minutes; an injection volume of 1 μL; column specifics: Agilent 19091S-433UI; splitless mode; an inlet temperature of 250 °C; a temperature range from -60 °C to 325 °C (maximum 350 °C); a total runtime of

53.5 minutes; Helium as the mobile phase at a flow rate of 1.2 mL/min; MS quad temperature at 150 °C; MS source temperature at 230 °C; a scan speed of 1,562 [N=2]; schema version: 2.3; electron ionization mode; mass range for molecular ions of 50-600 m/z; and peak area denoting relative percentages of constituents for the GC-MS analysis. The LC-MS analyses for the three ayurvedic preparations—Kottakkal Ayurveda Triphala (T), Kottakkal Ayurveda Hinguvachadi Churnam (H), and Kottakkal Ayurveda Jirakadyarishtam (J)—were executed using a 6550 iFunnel Q-TOF, adhering to standard procedures for LC-MS analysis (Hao, Lyu, Wang, & Xiao, 2023).

### 2.1.1. Evaluation of antimicrobial efficacy

Antimicrobial efficacy was assessed in vitro utilizing the disc diffusion technique.

### 2.1.2. Test Microorganism

Three Ayurvedic preparations, specifically Kottakkal Ayurveda Triphala (T), Kottakkal Ayurveda Hinguvachadi Churnam (H), and Kottakkal Ayurveda Jirakadyarishtam (J), were tested against the bacteria *Escherichia coli, Klebsiella pneumoniae, Staphylococcus aureus,* and *Streptococcus pneumoniae,* as well as the fungi *Aspergillus niger* and *Aspergillus fumigatus,* which were obtained from the National Chemical Laboratory (NCL) in Pune, India (Hassani, Durán, & Hacquard, 2018).

### 2.1.3. Preparation of Culture Media and Inoculum

Muller Hinton agar medium was composed of the following ingredients: Casein acid hydrolysate - 17.5g, Beef infusion extracts - 2.0g, Starch soluble - 1.5g, Agar - 17.0g, distilled water - 1000 ml, achieving a pH of 7.3 ± 0.1 at 25°C. A total of 34g of Muller Hinton agar was mixed in 1 liter of distilled water (Russell A. D., and J. R. Furr, 1977). The medium was pasteurized at 15 lbs of pressure at 121°C for 15 minutes. After the autoclaving process, the medium was poured into 100 mm Petri dishes and left to solidify. The nutrient broth liquid medium was composed of Peptone - 5.0g, Beef extract - 3.0g, NaCl - 5.0g, distilled water - 1000 ml, with a pH of 7.3 ± 0.1 at 25°C. The McFarland standard was readied by blending 0.05 ml of 1.175% $BaCl_2.H_2O$ with 9.95 ml of 1% $H_2SO_4$. Whatman No. 1 filter paper discs, each measuring 6 mm in diameter, were created and soaked with varying concentrations (25mg/ml, 50mg/ml, and 100mg/ml) of the three extracts, then air-dried at room temperature. Each disc was loaded with 30μg/ml of the corresponding extract (Leng, Hou, Xing, & Chen, 2024). The plates were incubated at 37 ± 2°C for 24 hours. Throughout the incubation period, the antimicrobial agent diffuses into the agar, preventing microbial proliferation. The plates were subsequently incubated at room temperature for an additional 2 to 4 days, permitting further diffusion and growth inhibition if the agent proved effective (Irobi, M. et al., 1994). The diameter of the zone of inhibition was gauged using a ruler, and the experiment was conducted in duplicate or triplicate.

### 2.1.4. In-vitro antidiabetic activity

### 2.1.5. Evaluation of In vitro Antidiabetic Activity through Glucose Uptake in a Yeast Cell Model

This model was previously described by Cirillo. 1% suspension of baker's yeast was created and permitted to stand at room temperature (25 °C) overnight, followed by centrifugation at 4,200 rpm for 5 minutes until a clear supernatant was achieved (Cirillo. V. P., 1962). Around 1–5 mg w/v of Kottakkal Ayurveda Triphala (T) extract was dissolved in DMSO until fully mixed. Different concentrations (5, 10, and 25 mM) of 1 mL simple sugar solution were then introduced into this mixture and brooded for 10 minutes at 37 °C. To commence the response, 100 μL of the species Saccharomyces was introduced to the glucose and extract combination, mixed gently, and further brooded for 60 minutes at 37 °C. After the brood period, the samples were seperated for 5 minutes at 3,800 rpm, and monosaccharide concentrations were assessed using a spectrophotometer (UV 5100B) at 520 nm. The percentage increase in glucose uptake was determined using the following formula: % increase in glucose uptake = (Absorbance of control - Absorbance of the sample) / Absorbance of control, where the control sample included all reagents except the test sample. Metronidazole was used as the standard reference drug.

### 2.1.6. Glucose Adsorption Assay

The extract's ability to adsorb glucose was evaluated according to the procedure outlined by Ou et al. Approximately 10 mg of Kottakkal Ayurveda Triphala (T) extract was mixed with 10 mL of simple sugar solution at three distinct concentrations (5, 10, and 20 mM). The mixtures were thoroughly blended and kept in a water bath shaker at 37 °C for 6 hours. After incubation, the samples were extracted at 4,800 rpm for 20 minutes, and the glucose levels in the supernatant were measured using a glucose oxidase assay kit. The quantity of glucose that was absorbed was calculated using the following formula:

**GLUCOSE BOUND = (G1 - G6) × volume of sample / Weight of the sample**

In this equation, G1 refers to the initial simple sugar concentration, whereas G6 denotes the simple sugar concentration after 6 hours (Ou, S., Kwok, K. C., Li, Y., and Fu, L., 2001).

### 2.1.7. Molecular docking investigation
### 2.1.8. Software and Tools

An investigation of molecular docking for the complex was carried out using Autodock, Schrodinger, BIOVIA Discovery Studio, and Desmond simulation software. The procedures for protein preparation, ligand processing, grid generation, docking, and results visualization were performed using Autodock Tools 4.2.6 and Discovery Studio Visualizer (Burley et al., 2021).

### 2.1.9. Docking Studies and Configuration of Protocols

The docking analysis entailed evaluating the phytoconstituents from the Ayurvedic formulations Kottakkal Ayurveda Triphala (T), Kottakkal Ayurveda Hinguvachadi Churnam (H), and Kottakkal Ayurveda Jirakadyarishtam (J) against the target protein 6ULM main protease and the human cadherin 17 EC1-2 receptors. The protocol for protein preparation was set by defining bond orders via the CCD database, while metal and disulfide bonds were constructed using a zero-order methodology (Schrödinger & DeLano, 2020). The Epik software produced different tautomeric states with a pH setting of N.N ±2.0. Following this, the 6ULM protein underwent minimization focusing on achieving a heavy atom RMSD of 0.30 Å and was analyzed through OPLS4 force field calculations. The docking preparation for the Ayurvedic phytoconstituents was executed using the LigPrep protocol, which employed OPLS4 force field parameters in addition to Epik, while also accounting for system desaltation. Tautomers and 32 stereoisomers were computed for each ligand. The receptor grid generation procedure considered the 6ULM-bound molecules, identifying that site as the binding pocket for the Ayurvedic phytoconstituents. The scaling factor for the Van der Waals radius was established at 1.0 with a cutoff for part charge set at 0.25. The pack surrounding the ligand was centered on the centroid of the workplace ligand, and the distance for the docked ligands was kept at 12 Å. A grid for docking the molecules was created based on the coordinates from the site map. The ligand docking rule was set with a calibrating aspect of 0.80 and a part charge stop of 0.15 for the van der Waals radii (Morris et al., 2009). The grid file was imported, and the XP extra precision option was employed alongside flexible ligand sampling. Considerations were made for nitrogen inversion sampling and rigid conformations, while all predefined functional groups were applied to direct the torsional sampling. Epik state fines were included in the final docking score (Wu, Liu, & Li, 2022). The energy window for ring sampling was fixed at 2.5 kcal/mol for generating conformers, and the distance-dependent dielectric constant for minimization was determined to be 2.0. The docking score evaluates specific poses by quantifying the strength of advantageous intermolecular interactions, such as hydrogen and hydrophobic bonds. Prime MM-GBSA provides the MMGBSA dG Bind equation as Complex − Receptor - Ligand, along with MMGBSA dG Bind (NS), yielding relative binding-free energies expressed in kcal/mol.

### 2.3.0. HCT 116 cell culture procedure

The HCT 116 cell line was adjusted to a concentration of $1\times10^4$ cells per well by utilizing DMEM media, along with an antimycotic antibiotic and 10% fetal bovine serum, in a 5% $CO_2$ brooder set to 37˚C.

### 2.3.1. Cytotoxic effects of Ayurvedic medicine on the HCT 116 cell line

Various concentrations of Ayurvedic medicine in SFM were prepared and brooded for a duration of 24 hours. Following the incubation period, the medium was discarded from the cells. A 5 mg/mL MTT solution in 1X PBS was then prepared and incubated for 4 hours at 37˚C using a $CO_2$ incubator. The resulting crystals were subsequently blended with 100 μL of DMSO, and the color intensity was measured at 570 nm. The absorbance was evaluated at 570 nm with a microplate reader, as the formazan dye transitions to a purple-blue hue (Shahid, Ramzan, Maurer, Parkman, & Fisher, 2012).

### 2.3.2. Cytotoxicity effects using the GES-1 cell line

The MTT assay is employed a favored method in academic laboratories, as evidenced by many articles published on this topic (Zhou, A. Y., Yuan, B., Liu, S. J., Wang, Y. F., & Li, S. S., 2021). The MTT substrate is prepared and then added to cultured cells, usually at various combinations of 5, 4, 3, 2, 1, 0.5, and 0.25 μg/mL, and incubated for a duration of 1 to 4 hours. Viable cells can convert MTT into a purple-colored formazan product when NADH donates electrons to MTT. In instances where cells experience death, they lose the ability to transform MTT into formazan, a process that is not fully understood; therefore, the resulting color formation serves as an effective marker for living cells only (Li & Yu, 2011). The formazan product from the MTT assay becomes concentrated as an insoluble precipitate within the cells, in addition to accumulating near the cell surface and in the culture medium (Yuan, B., Zhang, J. P., Wang, S. C., Shi, X. W., & Li, S. S., 2021).

Calculation of % Cell Viability = Sample OD × 100 / Control OD

Calculation of % Cell Cytotoxicity = 100 - Viability

### 3.1.1. Results and discussion

### 3.1.2. GC-MS evaluation of ayurvedic medicines

The Ayush department of the Indian government has advocated for the analysis of Ayurvedic medicines through GC-MS due to their international application. Research focusing on Ayurvedic medicines aimed to uncover chemical compounds, formulations, active ingredients, authenticate them, and assess their potential therapeutic qualities through molecular structure evaluations. GC-MS is a dependable analytical method for identifying and quantifying specific chemical constituents, which can bolster research and development for evidence-supported Ayurvedic treatments. Three Ayurvedic formulations, specifically Kottakkal Ayurveda Triphala (T), Kottakkal Ayurveda Hinguvachadi Churnam (H), and Kottakkal Ayurveda Jirakadyarishtam (J), underwent GC-MS examinations. The GC-MS chromatograms displayed 10, 15, and 8 peaks, suggesting the existence of 30, 45, and 8 chemical components in Kottakkal Ayurveda Triphala (T), Kottakkal Ayurveda Hinguvachadi Churnam (H), and Kottakkal Ayurveda Jirakadyarishtam (J), respectively. The first two Ayurvedic formulations, Kottakkal Ayurveda Triphala (T) and Kottakkal Ayurveda Hinguvachadi Churnam (H), produced 10 and 15 hits, with each yielding three biologically active substances per hit. The third formulation revealed a total of 8 phytochemical constituents. Each of these chemical components was acknowledged for their potential in addressing various health issues, as demonstrated by the GC-MS analysis. Numerous studies have highlighted that the three chosen Ayurvedic medicines can effectively manage multiple diseases, particularly hypertension, HIV, pain relief, cardiovascular problems, cancer, and diabetes. Information regarding the GC-MS results, including the nature, biological activity, hydrogen bond donors and acceptors, count of rotatable bonds, and heavy atom counts of the phytochemical constituents from Kottakkal Ayurveda Triphala (T), Kottakkal Ayurveda Hinguvachadi Churnam (H), and Kottakkal Jirakadyarishtam (J), can be found in **Tables 1, 2, and 3**. The chromatograms for Kottakkal Ayurveda Triphala (T), Kottakkal Ayurveda Hinguvachadi Churnam (H), and Kottakkal Ayurveda Jirakadyarishtam (J) are depicted in the accompanying **Figures 1a, 1b and 1c.**

### 3.1.3. Evaluation of Ayurvedic medicines using LC-MS

A thorough biochemical analysis is essential for Ayurvedic medicines to provide substantial evidence that aids in understanding the mechanisms by which certain chemical compounds may treat various ailments. The technique of LC-MS allowed for the identification of bioactive compounds, metabolites, mass spectra, and m/z values, ensuring both specificity and accuracy. The chromatogram obtained from LC-MS revealed multiple peaks that corresponded to the chemical compounds present in Ayurvedic preparations. Compound identification for Kottakkal Ayurveda Triphala (T), Kottakkal Ayurveda Hinguvachadi Churnam (H), and Kottakkal Ayurveda Jirakadyarishtam (J) was based on retention time, area, percentage area, and primarily m/z values of 15, 20, and 16, respectively. The LC-MS chromatograms for Kottakkal Ayurveda Triphala (T), Kottakkal Ayurveda Hinguvachadi Churnam (H), and Kottakkal Ayurveda Jirakadyarishtam (J) exhibited several peaks, which were counted as 15, 20, and 16 respectively. The LC-MS spectra for these Ayurvedic products—Kottakkal Ayurveda Triphala (T), Kottakkal Ayurveda Hinguvachadi Churnam (H), and Kottakkal Ayurveda Jirakadyarishtam (J)—are depicted in **Figures 2a, 2b and 2c**, respectively. Details regarding the retention time, area, and percentage area from the LC-MS spectra are shown in **Table 4.**

### 3.1.4. Spectroscopic investigations

Ayurvedic medicine does indeed provide effective treatment for existing health issues, which holds true. Modern spectroscopic techniques, including IR, UV, $^1$H-NMR, and LC-MS, were employed for the identification, quantification, and structural characterization of compounds found in or derived from botanical materials. These analytical approaches have demonstrated their usefulness in phytochemical and biological studies, aiding in the detection of biologically active compounds (Tangetal.,2019; Tansakul, Shibuya, Kushiro & Ebizuka, 2006; Kushiro, Shibuya & Ebizuka,1998; Han,Kim, Ban, Hwang,&Choi,2013).

### 3.1.5. Kottakkal Ayurveda Triphala (T)

Greyish-brown colour solid, IR (KBr): $\nu_{max}$ in cm$^{-1}$): 3855(O-H), 3439(–N-H), 2924(-COOH), 2853(-C-H), 2309(O=C=O), 2042(–C≡C–), 1632(C=O), 1497(C=C), 1384(C-N), 1236 (C=S), 1036 (C-S), 874(C-H) and 650 (C-Br); **(Figure 3a)** UV/Vis (CH$_3$OH): λmax (ε) = 840, 580 and 325 nm; **(Figure 3b)** $^1$H NMR (400 MHz, DMSO -d$_6$); δ: 1.22 (t, 3H, CH$_3$), 2.49 (m, 2H, CH$_2$), 3.34 (m, 2H, CH$_2$), 7.34-6.86 (m, 6H, Ar-H), 9.24 (s, 1H, -CHO); **(Figure 3c)** LCMS (m/z): 982, 981, 972, 958 and 933. **(Figure 3d)**

### 3.1.6. Kottakkal Ayurveda Hinguvachadi Churnam (H)

Brownish grey colour solid, IR (KBr): ν$_{max}$ in cm$^{-1}$): 3897(O-H), 3764 (O-H), 3431((–N-H), 2925(-C-H), 2855(-C-H), 2312(O=C=O), 2057(C=N), 1633(-COOH), 1383(C-N), 1238 (C–O–C), 1034 (–C=O), 874(C-H), 823(C=C), 605(C-S) and 537 (C-Br); **(Figure 4a)** UV/Vis (CH$_3$OH): λmax (ε) = 795, 580 and 330 nm; **(Figure 4b)** $^1$H NMR (400 MHz, CHCl$_3$ -D); δ: 1.25 (t, 3H, CH$_3$), 1.56 (s, 2H, CH$_2$) 2.09-2.01 (t, 3H, CH$_3$), 2.41-2.28 (q, 2H, CH$_2$), 3.90-3.80 (t, 2H, CH$_2$), 5.35 (s, 1H, -C=CH$_2$), 7.26 (s, 5H, C$_6$H$_5$); **(Figure 4c)** LCMS (m/z): 987, 985, 974 and 945. **(Figure 4d)**

### 3.1.7. Kottakkal Ayurveda Jirakadyarishtam (J)

Brown coloured liquid, IR (KBr): ν$_{max}$ in cm$^{-1}$): 3410(O-H), 2081(C≡C), 1638(C=O), 1411(C=C), 1224(C-O), 1056 (C-N), 599(C-Cl); **(Figure 5a)** UV/Vis (CH$_3$OH): λmax (ε) = 830, 450 and 340 nm; **(Figure 5b)** $^1$H NMR (400 MHz, CHCl$_3$ -D); δ: 1.25-1.22 (t, 3H, CH$_3$), 1.61 (s, 2H, CH$_2$) 3.74-3.69 (q, 2H, CH$_2$), 4.86 (s, 1H, OH), 7.26 (s, 5H, C$_6$H$_5$); **(Figure 5c)** LCMS (m/z): 969, 965, 951 and 941. **(Figure 5d)**

### 3.1.8. Results and discussion regarding molecular docking investigations

The results of the molecular docking investigations indicated that the bioactive compounds formed hydrogen bond interactions with amino acids including Phenylalanine, Isoleucine, Glutamine, Serine, Proline, Glutamic acid, Tyrosine, and L-lysine **(see Figure 6)**. The docking score values for Kottakkal Ayurveda Hinguvachadi Churnam (H) when assessed against the crystal structure of human cadherin 17 EC1-2 ranged from -4.494 kcal/mol to -1.535 kcal/mol (6ULM). The computed MMGBSA dG bind values for Kottakkal Ayurveda Hinguvachadi Churnam (H) were recorded between -41.15 kcal/mol and 10.91 kcal/mol. In a similar manner, Kottakkal Ayurveda Triphala (T) exhibited a docking score and an MMGBSA dG bind value against 6ULM that varied from -10.73 kcal/mol to 0.418 kcal/mol and -59.52 kcal/mol to -9.1 kcal/mol, respectively. On the other hand, Kottakkal Ayurveda Jirakadyarishtam (J) showed a docking score ranging from -3.815 kcal/mol to -1.572 kcal/mol, with MMGBSA dG bind values between -40.61 kcal/mol and -27.62 kcal/mol against the crystal structure of human cadherin 17 EC1-2 (6ULM). Kottakkal Ayurveda Triphala (T), Kottakkal Ayurveda Hinguvachadi Churnam (H), and Kottakkal Ayurveda Jirakadyarishtam (J) achieved superior docking scores of -4.494 (15Ha), -4.191 (3Hb), -4.031 (10Hc), -4.019 (15Hb), -4.004 (3Ha), -10.73 (8Ta), -9.108 (8Tc), -7.16 (8Tb), -5.208 (7Tc), -4.743 (3Tb), -4.122 (4Ta), -3.815 (5J), and -3.789 (6J). The MMGBSA dG bind values for Kottakkal Ayurveda Triphala (T), Kottakkal Ayurveda Hinguvachadi Churnam (H), and Kottakkal

Ayurveda Jirakadyarishtam (J) displayed higher readings of -42.97 (15Hb), -41.15 (10Hc), -40.14 (13Hc), -59.52 (8Ta), -53.87(1Tc), -55.8 (8Tc), -53.26 (8Tb), -41.21 (2J), and -40.61 (6J). The phytoconstituents 4Ha, 4Tc, and 4J did not interact with 6ULM main protease and the human cadherin 17 EC1-2 receptor, indicating that no conformations were generated. More favourable docking scores and MMGBSA dG bind values indicate stronger binding. The optimal binding configurations of the bioactive compounds from Ayurvedic medicines with 6ULM main protease and the human cadherin 17 EC1-2 receptor are depicted in the **Figure 7**. The molecular docking results corroborated the significant efficacy of these compounds against 6ULM main protease, human cadherin 17 EC1-2 receptor, and cancer. The **Table 5** presents the binding energies (Docking Score and MMGBSA dG Bind value) for different phytoconstituents in association with the 6ulm protein (human cadherin).

### 3.1.9. Microbial activity

Microorganisms like bacteria, fungi, and viruses can spread diseases to humans and animals through air, water, and food. Escherichia coli is a type of bacterium that affects the stomach and intestines, leading to symptoms such as stomach cramps, diarrhea, vomiting, and urinary tract infections. *Klebsiella pneumoniae* and *Streptococcus pneumoniae* cause various conditions, including sinusitis, pneumonia, osteomyelitis, arthritis, bacteremia, bloodstream infections, and meningitis. *Staphylococcus aureus* is a flexible pathogen capable of inducing a range of illnesses, from mild to severe. The *fungi Aspergillus niger* and *Aspergillus fumigatus* are linked with infectious diseases that include pneumonia, lung disorders, chronic lung infections, bronchopulmonary diseases, thrombosis, and allergic reactions. Historically, the following medications (Kottakkal Ayurveda Triphala (T), Kottakkal Ayurveda Hinguvachadi Churnam (H), and Kottakkal Ayurveda Jirakadyarishtam) have been recommended for these ailments. Therefore, these preparations were selected to evaluate their potential antimicrobial effects against various pathogenic microorganisms.

The antibacterial properties of Kottakkal Ayurveda Triphala (T) were assessed against both Gram-positive (*Streptococcus pneumoniae* and *Staphylococcus aureus*) and Gram-negative bacteria (*Escherichia coli* and *Klebsiella pneumoniae*) using the Disc-Diffusion method on Muller Hinton Agar. Kottakkal Ayurveda Hinguvachadi Churnam (H) and Kottakkal Ayurveda Jirakadyarishtam (J) were tested against fungi (*Aspergillus niger* and *Aspergillus fumigatus*) utilizing the same agar method. Synthetic antibiotics *Ofloxacin* and *Amoxicillin* with clavulanic acid served as control comparisons. An antimicrobial testing solution was prepared at a concentration of 30μg/ml. The polyherbal Ayurvedic treatments were tested at three different concentrations: 25, 50, and 100 mg/ml. Distinct media were employed for bacteria and fungi, specifically nutrient agar and Sabouraud dextrose agar. Both types of media were incubated at 37°C for a 24-hour duration to promote microbial growth. Plates treated with the solutions were then placed on the media and inoculated with Gram-positive (*Streptococcus pneumoniae* and *Staphylococcus aureus*) and Gram-negative (*Escherichia coli* and *Klebsiella pneumoniae*) bacterial strains, as well as fungal strains (*Aspergillus niger* and *Aspergillus fumigatus*). The zones of inhibition were measured, and the results were recorded as the diameter of the inhibition zones in millimeters. The results of the antibacterial and antifungal assays are summarized in **Table 6 and Figures 8 and 9.**

Kottakkal Ayurveda Triphala (T) exhibited notable inhibition, with measurements of 16, 14, 9, and 13 mm at a concentration of 100 mg/ml against *Escherichia coli, Klebsiella pneumoniae, Staphylococcus aureus, and Streptococcus pneumoniae,* respectively, when compared to the standard antibiotics *Ofloxacin* and *Amoxicillin* with clavulanic acid, which presented inhibition zones of 17, 16, 13, and 16 mm at 30 μg/ml. The remarkable antimicrobial properties of this formulation are attributed to a range of phytoconstituents and bioactive compounds such as Lincomycin, 2-Acetyl-2,3,5,6-tetrahydro-1,4-thiazine, 4H-Pyran-4-one, 2,3-dihydro-3,5-dihydroxy-6-methyl-, 2,4-Dihydroxy-2,5-dimethyl-3(2H)-furan-3-one, Catechol, 1H-Pyrazole, 4,5-dihydro-3-methyl-1-propyl-, 1-Ethyl-2-hydroxymethylimidazole, 1,2,4-Benzenetriol, Methylparaben, Benzoic acid, 3-hydroxy-, methyl ester, and Isopropyl-4-hydroxybenzoate present in Kottakkal Ayurveda Triphala (T).

Inhibition zones measuring 15 and 6 mm were noted against *Aspergillus niger*, while for *Aspergillus fumigatus,* inhibition zones of 14 and 16 mm were observed at the same concentration of 100 mg/ml. The antifungal controls demonstrated inhibition measurements of 16 and 17 mm at a concentration of 30 μg/ml. The antimicrobial activity was evaluated through the ANOVA method, showing significance at P=0.0650. The antifungal properties of Kottakkal Ayurveda Hinguvachadi Churnam (H) and Kottakkal Ayurveda Jirakadyarishtam (J) showed outcomes like those of the standard treatments. The notable antifungal potency of these two herbal formulations can be attributed to phytoconstituents such as Tetramethyl phosphonium cation, 8,9-Dehydro thymol, Thymol, 3-Methyl-4-isopropyl phenol, Bicyclo[5.2.0]nonane, 2-methylene-4,8,8-trimethyl-4-vinyl-, Benzene, 1-(1,5-dimethyl-4-hexenyl)-4-methyl-, Ethyl p-methoxycinnamate, Dimoxystrobin, Hexadecanoic acid, 2-hydroxy-1-(hydroxymethyl)ethyl ester, Glycerol 1-palmitate, n-Propyl 9-tetradecenoate, and Diethyl Phthalate.

## 3.2.0. In vitro antidiabetic properties

### 3.2.1. Yeast Cell Test for Glucose Absorption

High blood glucose levels, associated with diabetes, can negatively impact the eyes, nerves, heart, and kidneys; sustained elevated glucose can also contribute to cancer development. The most common types of diabetes include type 1, type 2, and gestational diabetes. This study investigated the antidiabetic effects of Kottakkal Ayurveda Triphala (T) using in vitro techniques, particularly through alpha-amylase and glucose uptake tests with a yeast model. In this experiment, yeast cells were mixed with three different glucose concentrations (5, 10, and 20 mM) along with five distinct amounts of Kottakkal Ayurveda Triphala (T), specifically 1, 2, 3, 4, and 5 mg (Glucose + Yeast + Ayurvedic medicine), in addition to a control group (Glucose + Yeast) and a standard dose of Metformin (5 mg) to evaluate the in vitro antidiabetic effect. The optical density (OD) values indicating glucose absorption by yeast cells are detailed in the **tables 7and 8** and **figures 10 and 11** (Zhou, Yuan, Liu, Wang, & Li, 2021).

After treating yeast cells with Kottakkal Ayurveda Triphala (T), an increase in the percent of glucose uptake by yeast cells was noted across different glucose concentrations (5, 10, and 20 mM) when compared to the Ayurvedic medicine's concentrations of 1, 2, 3, 4, and 5 mg/ml. The observed glucose uptake values were 62.21 (5mM), 71.78 (5mM), 72.11 (5mM), 72.98 (5mM), 73.52 (5mM), and 81.48 (5mM), in addition to 42.19 (10mM), 48.67 (10mM), 51.91 (10mM), 52.01 (10mM), 58.19 (10mM), and 60.84 (10mM), as well as 13.43 (20mM), 17.73 (20mM), 21.50 (20mM), 24.45 (20mM), 25.13 (20mM), and 33.50 (20mM) of glucose. The highest concentration of Kottakkal Ayurveda Triphala (T) showed the most activity at all glucose levels, reaching values of 73.52 and 81.48 for (5mM). Each concentration of Kottakkal Ayurveda Triphala (T) exhibited adsorption, with the binding capacity directly proportional to the molar concentration of glucose (Yuan, Zhang, Wang, Shi, & Li, 2021). A simple glucose adsorption assay was conducted to measure the amount of bound glucose, which was determined to be 98.5% [G1-1.312 & G2-0.998; Sample Weight – 1mg, Sample Volume – 100ml; Concentration = 20mM], 82% [G1-1.021 & G6-0.821; Sample Weight – 1mg; Sample Volume – 100ml; Concentration = 10mM], and 71.4% [G1-0.921 & G6-0.723; Sample Weight – 1mg; Sample Volume – 100ml; Concentration = 5mM]. In-vitro antidiabetic activity was assessed using the ANOVA method, demonstrating significance at $P<0.0001$ (Tan, Wang, Zhou, & Si, 2021). The data suggest that higher concentrations of Kottakkal Ayurveda Triphala (T) are associated with an increased percentage of glucose uptake by yeast cells. Kottakkal Ayurveda Triphala (T) seems to have notable antidiabetic properties attributed to the presence of antidiabetic phytoconstituents like 2-Acetyl-2,3,5,6-tetrahydro-1.

### 3.2.3. Anticancer research

Various treatments from Ayurvedic, Siddha, and homeopathic systems are recognized for their anticancer properties, as shown in multiple studies. Global research efforts are underway to discover effective therapies aimed at curing cancer in humans. The fields of Ayurveda, Siddha, Unani, and Homeopathy have long been dedicated to this objective. Phytochemicals have gained significant interest for their various biological effects, including cytotoxic and chemopreventive properties. Three Ayurvedic formulations—Kottakkal Ayurveda Triphala (T), Kottakkal Ayurveda Hinguvachadi Churnam (H), and Kottakkal Ayurveda Jirakadyarishtam (J)—were selected for assessment due to their traditional and documented uses in treating early-stage colorectal cancer. This study examines the cytotoxic effects of the three Ayurvedic treatments on the human colorectal carcinoma cell line (HCT 116) in vitro. Extracts from the three Ayurvedic medicines at concentrations of 5, 4, 3, 2, 1, 0.5, and 0.25 (µg/ml) were prepared and evaluated for their impact on the in vitro human colorectal carcinoma cell line (HCT 116) at various concentrations. The survival rates of the cultured human colorectal carcinoma cells (HCT 116) subjected to the Ayurvedic extracts were monitored closely. The extracts of Kottakkal Ayurveda Triphala (T), Kottakkal Ayurveda Hinguvachadi Churnam (H), and Kottakkal Ayurveda Jirakadyarishtam (J) at a concentration of 5 (µg/ml) inhibited the growth of the human colorectal carcinoma cell line (HCT 116) by 33.42%, 28.06%, and 26.79%, respectively. The distinct anticancer effects of these plant extracts were evaluated. These three formulations merely decreased the proliferation of cancer cells by 27 to 33% (Bi et al., 2021; Ma & Wang, 2021; Weiet al., 2021).

The results on cytotoxicity are shown as the percentage of cancer cell death caused by the ayurvedic extracts, compared to control cells treated with DMSO/PBS at the same concentration as the extracts, as depicted in the accompanying **Table 9 and figures 12, 13 and 14.** The data reveal that the three ayurvedic extracts effectively inhibited cancer cell growth at lower concentrations, with varying degrees of efficacy depending on the specific extract and the cancer cell line assessed. The experiment indicated that the anticancer effects were influenced by both time and concentration, and importantly, did not harm healthy human cells. These results demonstrated that the extracts from the three ayurvedic medicines at concentrations of 5, 4, 3, 2, and 1 (µg/ml) induced cancer cell death through different mechanisms via their bioactive compounds. The three products studied, Kottakkal Ayurveda Triphala (T), Kottakkal Ayurveda Hinguvachadi Churnam (H), and Kottakkal Ayurveda Jirakadyarishtam (J), were evaluated using in vitro cancer models and displayed considerable inhibition of cancer cell proliferation. Kottakkal Triphala (T) encompasses bioactive phytochemicals such as 1Ta, 2Ta, 2Tb, 3Ta, 4Ta, 4Tb, 5Tb, 7Ta, and 9Tb, which are known for their antioxidant and anticancer characteristics. Kottakkal Ayurveda Hinguvachadi Churnam (H) contains bioactive substances like 2Hc, 4Ha, 4Hb, 5Hb, 5Hc, 7Ha, 7Hb, 8Ha, 8Hb, 8Hc, 9Ha, 10Ha, 10Hb, 10Hc, 11Ha, 11Hb, 12Ha, 13Hb, 13Hc, 14Hc, 15Ha, 15Hb, and 15Hc, which are likewise recognized for their antioxidant and anticancer effects. The specific pharmacological compounds 2J, 5J, 7J, and 8J found in Kottakkal Ayurveda Hinguvachadi Churnam (H) are acknowledged for their antioxidant and anticancer properties.

## 4.Conclusion

This report provides the initial exploration employing GC-MS, LC-MS, spectroscopic techniques, and docking analyses of the Ayurvedic remedies Kottakkal Ayurveda Triphala (T), Kottakkal Ayurveda Hinguvachadi Churnam (H), and Kottakkal Ayurveda Jirakadyarishtam (J), alongside their assessments for antimicrobial, antidiabetic, and cytotoxic activities. The structures and functional groups of the phytoconstituents were characterized using GC-MS, LC-MS, and spectroscopic techniques. The analyzed Ayurvedic preparations are rich in flavonoids, steroids, alkaloids, quinolone, curcumin, cuparene, organophosphorus compounds, terpenes, quinol, pyrogallol, Tylenol, and various aldehydes, amides, ketones, esters, aromatic compounds, and alcohols. The three Ayurvedic formulations exhibited cytotoxic effects in vitro against a human colorectal carcinoma cell line (HCT 116), with cancer cell inhibition rates ranging from 27% to 33%. The investigation assessed the antidiabetic efficacy of Kottakkal Ayurveda Triphala (T) through a glucose uptake assay utilizing yeast cells, indicating that it can significantly reduce the risk of diabetes by assisting in the digestion of food in the stomach. Additional antimicrobial evaluations were performed, with Kottakkal Ayurveda Triphala (T) tested against Escherichia coli, Klebsiella pneumoniae, Staphylococcus aureus, and Streptococcus pneumoniae, while Kottakkal Ayurveda Hinguvachadi Churnam (H) and Kottakkal Ayurveda Jirakadyarishtam (J) were examined against Aspergillus niger and Aspergillus fumigatus. The results suggest that all three Ayurvedic remedies exhibited significant antimicrobial properties against the tested microorganisms, which can be attributed to the discovery of novel biologically active compounds. These Ayurvedic formulations seem capable of inhibiting or averting diseases related to microbes, diabetes, and intestinal cancer. Further research on the chosen Ayurvedic remedies requires additional methods and effort.

**Disputes of interest**

**Financial assistance and support**

It is stated that the authors' study was conducted without any funding or sponsorship.

**Acknowledgement**

For their unwavering support and encouragement, the authors would like to thank the Chairman, Honourable Secretary & Correspondent, Principal, Vice-Principals (Academic & Administration), and Head, P.G. & Research Department of Chemistry, The New College (Autonomous), Chennai.